# Effects of Fe doping in $La_{1/2}Ca_{1/2}MnO_3$

P. Levy, L. Granja, E. Indelicato, D. Vega, G. Polla and F. Parisi

Grupo Materia Condensada, Departamento de Física, Centro Atómico Constituyentes,
Comisión Nacional de Energía Atómica, Av. Gral Paz 1499 (1650) San Martín, Argentina.

*The effect of Fe doping in the Mn site on the magnetic, transport and structural properties of polycrystalline $La_{1/2}Ca_{1/2}MnO_3$ was studied. Doping with low Fe concentration ( < 10%) strongly affects electrical transport and magnetization. Long range charge order is disrupted even for the lowest doping level studied (~2%). For Fe concentration up to 5% a ferromagnetic state develops at low temperature with metallic like conduction and thermal hysteresis. In this range, the Curie temperature decreases monotonously as a function of Fe doping. Insulating behavior and a sudden depression of the ferromagnetic state is observed by further Fe doping.*

Mixed valence manganates $Ln_{1-x}D_xMnO_{3+\delta}$ (where Ln is a lanthanide and D an alkaline earth ion) have been the subject of broad research because of the variety of properties they exhibit [1]. Among them, those compounds exhibiting charge order (CO) phenomena are being the focus of intense studies because percolative paths of coexisting ferromagnetic (FM) metallic phases are suspected to be responsible of colossal magnetoresistance[2]. The low temperature ordering of $Mn^{+3}$ and $Mn^{+4}$ can be destroyed by substitutionally induced cation disorder, either in the Mn site[3] or in the Ln-D site[4]. Also, application of magnetic field may lead to a charge melting [5] of the otherwise ordered lattice.

The $La_{1/2}Ca_{1/2}MnO_3$ compound has antiferromagnetic (AFM) ground state with CO; upon warming, a first order phase transition to a FM state is observed. Thermal disorder inhibits charge localization and AFM coupling, and thereby the double exchange (DE) mechanism leads to a FM state. The subtle competition between FM and CO states gives rise to phase separation into nanodomains of FM and CO states below 150K [6]. We have studied the effect of Fe substitution in the Mn site of $La_{1/2}Ca_{1/2}MnO_3$. Low temperature long range CO is disrupted and a FM state is obtained for Fe concentration below 5%. By further Fe doping insulating behavior and a sudden depression of the ferromagnetic state is observed.

Polycrystalline samples of $La_{1/2}Ca_{1/2}Fe_yMn_{1-y}O_3$ (0 < y < 0.1) were prepared by thermal decomposition of the corresponding citrates at 700 °C, followed by a heat treatment at 1000 °C for 10 hours. Powders were pelletized and sintered at the same temperature for 5 hours. X-ray powder diffraction data was collected at room temperature using a Philips diffractometer. Magnetization was measured using a commercial SQUID magnetometer. Resistivity was measured by the four probe technique.

The obtained samples were single phase and all reflections could be indexed on the basis of an orthorhombic cell with space group *Pnma* and lattice parameters a = 5.4148(5) Å, b = 7.6389(7) Å, c = 5.4260(5) Å for the undoped sample. A slight and smooth increase in lattice parameters as a function of y was obtained, of the order of 0.3% for y = 0.1. Mössbauer spectra confirmed the presence of the dopant as $Fe^{+3}$.

The temperature dependence of the magnetization at H = 1 T is shown in Figure 1 for samples with different Fe doping level. The undoped sample displays the typical behavior already reported in the literature: paramagnetic above 250 K, FM between $T_C$ = 225 K and 150 K on cooling, and AFM below 150 K. Low doping level produces a gradual decrease in the Curie temperature, and an abrupt increase of the low temperature magnetization up to 3 $\mu_B$ for y = 0.05. Beyond 5% Fe doping a sudden decrease in the low temperature magnetization is observed, obtaining 0.3 $\mu_B$ for y = 0.07.

The effect of Fe doping on $\rho(T)$ at H=0 is shown in figure 2 The undoped sample displays thermally activated resistivity behavior above and close below $T_C$ = 225 K. At T = 150 K on cooling (200 K on warming) the resistivity increases (decreases) abruptly due to the CO transition. This low temperature CO phase present in the undoped sample is inhibited for y = 0.02, forcing a more than 7 orders of magnitude change in the resistivity value. Metallic like behavior is observed below 120 K on cooling, with a resistivity peak at 140 K. Above 200 K $\rho(T)$ follows activated behavior. Thermal hysteresis is observed: $\rho(T)$ around the peak is higher for cooling curves than for those obtained upon warming the sample from 30 K. The temperature range where hysteresis occurs is about 80 K. Interestingly, the nature of the irreversibility is qualitatively different to that depicted by the undoped sample, reflecting a change in the ground state. Samples with doping levels y = 0.03 and 0.05 display a similar features to the y = 0.02 sample described above, with increased resistivity values and decreased peak resistivity temperatures. For y = 0.07 insulating reversible behavior is obtained down to 30 K, although low magnetization and depressed $T_C$ can still be observed in magnetization data.

The effect of Fe on the $La_{1/2}Ca_{1/2}Fe_yMn_{1-y}O_3$ compound can be analyzed from two different points of view, namely the effects on the CO phase and on the DE coupling. The undoped compound is now known to exhibit phase separation into nanodomains[6] of FM and CO states below 150 K. In our case, the low temperature magnetization of 0.3 $\mu_B$ at H = 1 T calls for about a 10 % volume fraction of FM phase imbedded in a CO insulating phase with no metallic percolation through the sample. As low Fe doping levels are introduced, low temperature long range CO is inhibited and metallic percolation is induced through the whole sample. Magnetic measurements show that concomitantly a robust low temperature FM phase develops. Thus it is suspected that low Fe doping stabilizes enough FM metallic clusters to account for both experimental facts.

As Fe does not participate in the DE mechanism, it weakens the FM phase when it is introduced in the Mn site. Quantitative considerations suggest a non local effect, instead of a single Mn-O-Mn bond interruption produced by each Fe ion introduced. In the related compound with pure FM metallic ground state ($La_{2/3}Ca_{1/3}MnO_3$) a decrease in $T_C$ and low field magnetization is obtained [7] when Fe is doping the Mn site. The increase in resistivity values in the FM phase of lightly doped compounds and the monotonous decrease in $T_C$ as Fe doping increases can then be related to the weakening of the DE mechanism within FM clusters.

Summarizing, we have observed suppression of long range CO in $La_{1/2}Ca_{1/2}MnO_3$ by low level doping with Fe the Mn site. At low temperature a FM state develops, which is further depressed by the increase of the impurity concentration. Phase separation into FM and CO states in the undoped compound, the presence of thermal hysteresis and enhanced magnetoresistance on Fe doped samples suggest that CO regions coexist with percolative FM regions.

Acknowledgments: We thank L.Civale for help during magnetization measurements. Part of this work was supported by CONICET PEI 98 #125.

**Figure Captions**

Figure 1: Field cool cooling magnetization as a function of temperature at H=1T for $La_{1/2}Ca_{1/2}Fe_yMn_{1-y}O_3$ (0 < y < 0.07) samples.

Figure 2: Temperature dependence of resistivity for $La_{1/2}Ca_{1/2}Fe_yMn_{1-y}O_3$ (0 < y < 0.07) samples. Arrows indicate the directions of the temperature change.

Figure 1

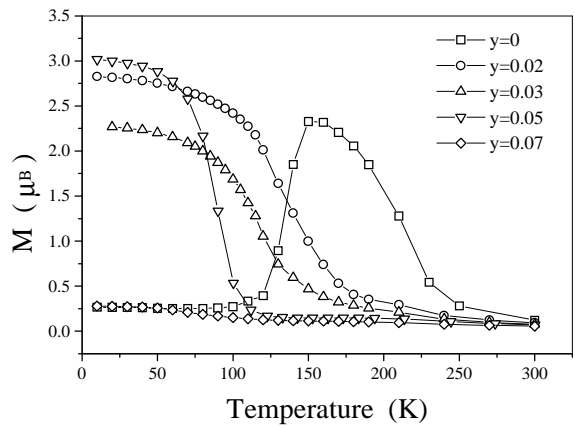

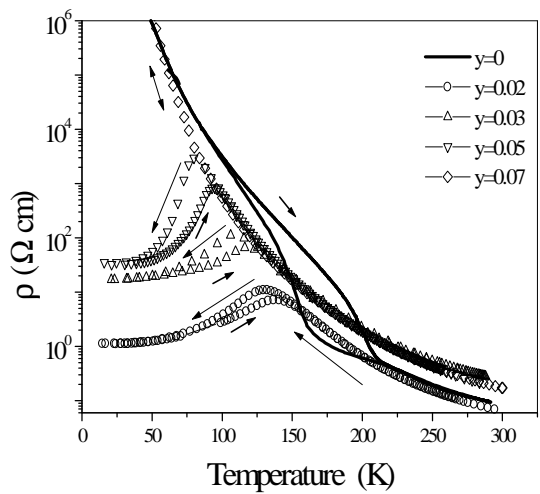

Figure 2